\documentclass[pra,aps,twocolumn,showpacs]{revtex4}

\usepackage{amsmath}
\usepackage{amssymb}
\usepackage{graphicx}
\usepackage{hyperref}
\usepackage{verbatim}

\newcommand{\cel}{\,^\circ\mathrm{C}}
\newcommand{\ee}[1]{\times 10^{#1}}
\newcommand{\unit}[1]{\,\mathrm{#1}}

\newcommand{\Rb}{^{87}\mathrm{Rb}}

\newcommand{\Plug}{P_{\rm plug}}

\usepackage{times}

\begin{document}

\title{A compact single-chamber apparatus for Bose-Einstein condensation of $^{87}$Rb}
\author{Igor Gotlibovych, Tobias F. Schmidutz, Stuart Moulder, Robert L. D. Campbell, Naaman Tammuz, Richard J. Fletcher, Alexander L. Gaunt, Scott Beattie, Robert P. Smith, and Zoran Hadzibabic}
\affiliation{Cavendish Laboratory, University of Cambridge, J. J. Thomson Avenue, Cambridge CB3 0HE, United Kingdom}

\date{\today}

\begin{abstract}
We describe a simple and compact single-chamber apparatus for robust production of $\Rb$ Bose-Einstein condensates. 
The apparatus is built from off-the-shelf components and allows production of quasi-pure condensates of $>3\times10^5$ atoms in $<30\;$s. This is achieved using a hybrid trap created by a quadrupole magnetic field and a single red-detuned laser beam [Y.-J. Lin {\it et~al.}, Phys. Rev. A {\bf 79},  063631  (2009)]. In the same apparatus we also achieve condensation in an optically plugged quadrupole trap [K.~B. Davis {\it et~al.}, Phys. Rev. Lett. {\bf 75},  3969  (1995)]; we show that as little as $70\;$mW of plug-laser power is sufficient for condensation, making it viable to pursue this approach using inexpensive diode lasers. 
While very compact, our apparatus features sufficient optical access for complex experiments, and we have recently used it to demonstrate condensation in a uniform optical-box potential [A. Gaunt {\it et al.}, arXiv:1212.4453 (2012)].
\end{abstract}



\maketitle

\section{Introduction}\label{sec:intro}
Atomic Bose-Einstein condensates (BECs) are now widely used for studies of many-body physics~\cite{Bloch:2008} and as bright sources for atom optics and interferometry~\cite{Cronin:2009}. Since the first demonstrations of condensation in magnetic \cite{Anderson1995, Bradley1995, Davis1995} and optical \cite{Barrett2001} traps, the development of experimental setups for BEC production has generally taken two parallel routes. On one hand, some applications require ever more complex setups; these, for example, include multi-species experiments~\cite{Tablieber:2008}, single-particle detection~\cite{Bakr:2009} or hybrid systems combining neutral atoms and ions~\cite{Zipkes:2010}.
On the other hand, there has also been a continuing effort to design simple and inexpensive machines that are still suitable for many experiments \cite{Lewandowski2003,Du2004, Naik2005, Heo2011,Weber2003, Kinoshita2005, Jacob2011, Arnold2011a}.

\begin{figure}[bt]
\begin{center}
\includegraphics[width=\linewidth]{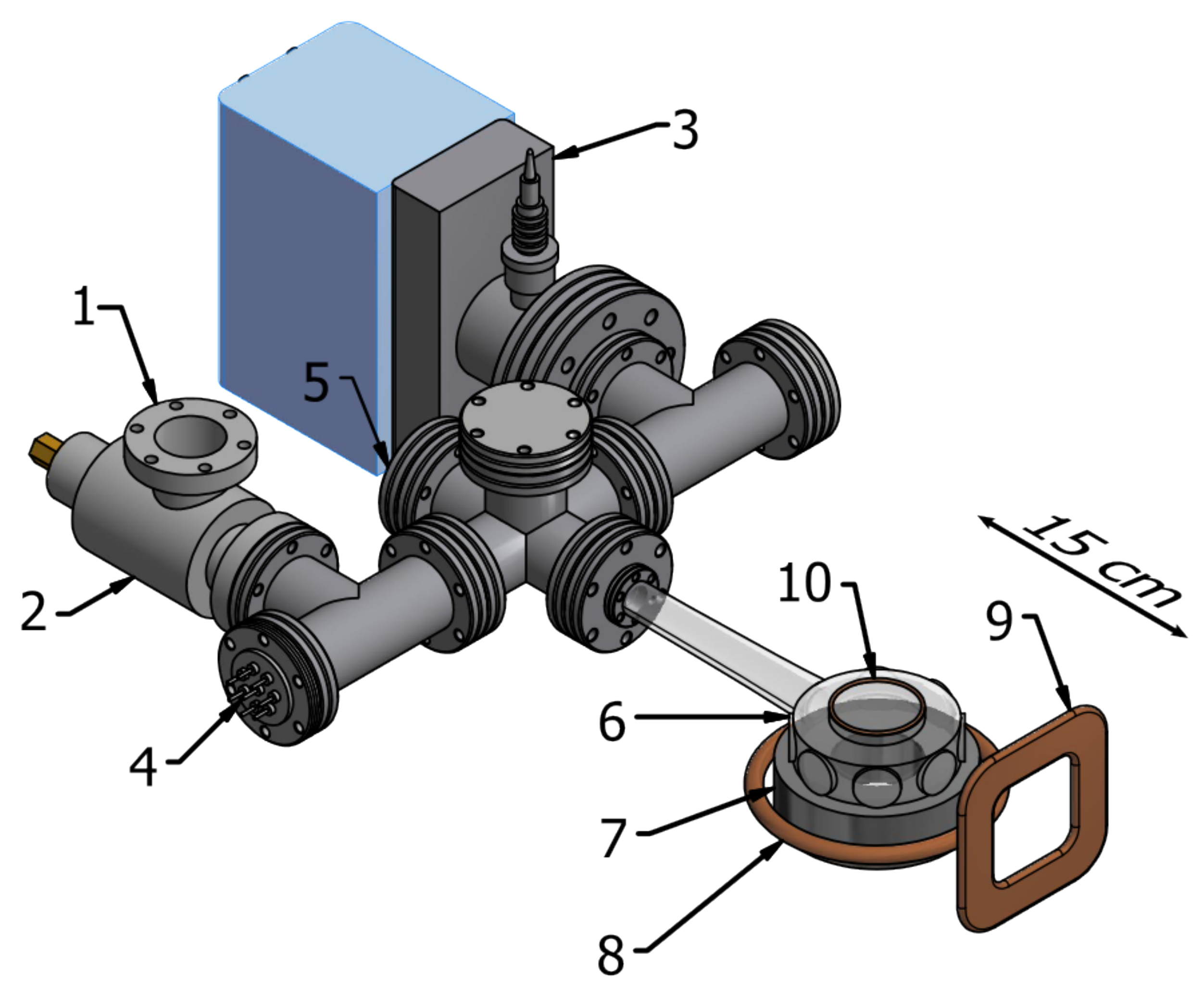}
\end{center}
\caption{Vacuum system: 1 - turbo pump port, 2 - valve, 3 - ion pump, 4 - $\Rb$ dispensers, 5 - viewport, 6 - glass cell. We also show the magnetic coils for creating: 7 - quadrupole field, 8 - bias field, 9 - guide field, 10 - RF field. All coils except 9 and 10 are paired (only one coil per pair is shown).}
\label{fig:vacuum}
\end{figure}

In this paper, we describe a particularly simple apparatus, built exclusively from off-the-shelf components, and our cooling procedure for production of $\Rb$ BECs. 
While being compact and consisting of a single vacuum chamber for both laser and evaporative cooling (see Fig.~\ref{fig:vacuum}), our setup features BEC atom numbers and production rates comparable to most multi-chamber $\Rb$ setups, as well as sufficient optical access for complex experiments.
By evaporative cooling in a hybrid magnetic-optical trap introduced by Lin {\it et al.}~\cite{Lin2009}, we reliably produce quasi-pure BECs of $>3\times 10^5$ atoms in $<30\;$s.  
Using the same setup, we also explored condensation in an optically plugged quadrupole trap~\cite{Davis1995}. Specifically, we study the plugging laser power requirements for a range of plug-laser wavelengths and show that it is possible to achieve condensation using as little as $70\;$mW of plug power. This opens the possibility of implementing this approach using inexpensive diode lasers.
Recently, we have used this apparatus (in the hybrid-trap configuration) to demonstrate Bose-Einstein condensation in a uniform optical-box potential~\cite{Gaunt:2013}. This unambiguously shows that this simple and inexpensive setup is sufficient for many non-trivial experiments requiring significant optical access to the atomic cloud.

The article is structured as follows. In Sec.~\ref{sec:setup} we outline our experimental setup. In Sec.~\ref{sec:MOT} we provide details of the laser cooling stage of our experiments. Evaporative cooling and condensation in a hybrid trap are described in Sec.~\ref{sec:Hybrid}, while experiments with an optically plugged trap are discussed in  Sec.~\ref{sec:plug}. Finally, in Sec.~\ref{sec:conclusion} we summarise our results.

\section{Experimental Setup}\label{sec:setup}
\subsection{Vacuum System}

BEC production in a single-chamber vacuum system is subject to two contradicting requirements: while a higher $\Rb$ vapour pressure facilitates loading of the magneto-optical trap (MOT), a lower background gas pressure increases the lifetime and evaporation efficiency during later cooling stages. Condensation is achieved by carefully balancing these two requirements. In our experiments this balance is improved by using isotopically pure $\Rb$ vapour sources (purchased from Alvatec).

Our vacuum system is shown in Fig.~\ref{fig:vacuum}. A quartz cell by Triad Technology allows optical access via  two $50\unit{mm}$ (top and bottom) and eight $25\unit{mm}$ anti-reflection coated windows. The vacuum is maintained by a  single $45 \unit{l/s}$ ion pump. 
After moderate baking of the system, at $200\cel$ for a week, we achieve magnetic trap lifetimes of over $20\unit{s}$. The $\Rb$ pressure typically reduces this to about $10\unit{s}$.

\subsection{Magnetic Coils}

The magnetic quadrupole field for both the MOT and the magnetic trap is created by an anti-Helmholtz pair of 16-turn water-cooled coils wound from 4~mm copper tubing (``7" in Fig.~\ref{fig:vacuum}). These create an axial gradient of $B' = 400\unit{G/cm}$ when carrying $200\unit{A}$.

The quantisation axis for optical pumping and imaging is provided by a magnetic field along the imaging axis. We use a low-inductance planar 10 turn coil wound from $5\unit{mm}$ copper wire (``9" in Fig.~\ref{fig:vacuum}), providing a field of $5\unit{G}$ at the atoms.

Gravity compensation during time-of-flight (TOF) measurements is achieved by combining a quadrupole field (created by the MOT coils) with a homogeneous vertical bias field. Residual potential curvature is minimised by a high bias field strength. We use two 50-turn coils in a Helmholtz configuration (``8" in Fig.~\ref{fig:vacuum}), providing a bias field of $70\unit{G}$ when run at 10~A.

A radio-frequency (RF) field for forced evaporative cooling is created by a three-turn coil mounted against the top viewport of the cell.
For transitions between neighbouring $m_F$ states separated by up to $15\unit{MHz}$, we achieve Rabi frequencies above $20\unit{kHz}$ using $600\unit{mW}$ of RF power.

\subsection{Laser-Cooling System}

Magneto-optical trapping of $\Rb$ requires two laser frequencies of its $\textrm{D}_2$ line (near $780\unit{nm})$. We use two grating-stabilised external-cavity diode lasers (DL Pro from Toptica Photonics). They are referenced to separate vapor cells to generate cooling light close to the $|F=2\rangle \rightarrow |F'=3\rangle$ cycling transition and  repumping light resonant with the $|F=1\rangle \rightarrow |F'=2\rangle$ transition. Cooling light is amplified using a tapered amplifier (BoosTA from Toptica). The cooling and repumping light pass through acousto-optic modulators (AOMs) and are combined in a two- to six-way free-space fiber port cluster, built in-house from OFR components (distributed by Thorlabs).

\begin{figure}[t]
\begin{center}
\includegraphics[width=\columnwidth]{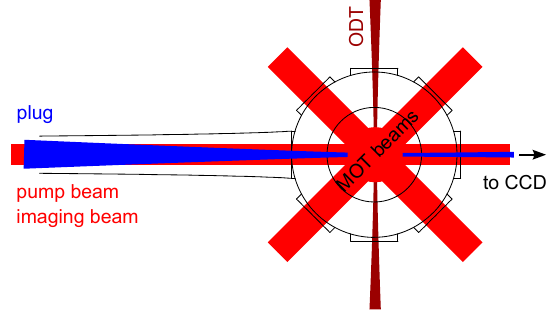}
\end{center}
\caption{Top view of the glass cell, showing the arrangement of the laser beams.}
\label{fig:beams}
\end{figure}

The spatial arrangement of our laser beams is shown in Fig.~\ref{fig:beams}.
The six MOT beams are delivered to the glass cell via optical fibers. A total of $120\unit{mW}$ of cooling light and $10\unit{mW}$ of repumping light reaches the atoms. We use large diameter ($25\unit{mm}$) beams which do not need to be realigned for months.
The light for optical pumping and absorption imaging of the atoms is derived from the cooling light. The pumping and imaging beams are controlled by additional AOMs and enter the chamber via the ``back" viewport (5 in Fig.~\ref{fig:vacuum}).

\subsection{Single-Beam Dipole Trap}

The optical-dipole-trap (ODT) beam used for the hybrid trap is derived from a $20\unit{W}$ ytterbium fiber laser  (YLR-20-LP from IPG Photonics). After passing through an AOM, the beam is focused by a concave-convex pair of lenses, allowing the adjustment of both waist size and position. We typically use a beam waist of $65\unit{\mu m}$ and a maximum power of $5\unit{W}$ at the atoms. Higher powers lead to excessive atom losses, which can be attributed to two-photon transitions~\cite{Lauber2011b}; these transitions are caused by the large linewidth of the trapping laser ($1.8 \unit{nm}$ in our case). The focal point is positioned approximately a beam waist below the zero of the magnetic quadrupole field. The beam power is controlled in the range $0.01-5\unit{W}$ using a feedback loop consisting of a photodiode, a proportional-integral-derivative (PID) controller, and the AOM. Throughout this range, we achieve an rms power noise below $1\unit{mW}$ over a bandwidth of $1\unit{kHz}$.

\subsection{Optical-Plug Beam}

In our experiments, for the optical-plug light we used a Ti:Sapphire laser (Coherent MBR-110) in order to explore a range of plug wavelengths and powers. 
However, since we show that as little as $70\unit{mW}$ of plug power is sufficient to achieve quantum degeneracy, in future experiments the plug light could be generated using a simple diode laser. 

The plug beam is focused to a waist of $27\unit{\mu m}$ and aligned to overlap with the zero of the magnetic quadrupole field, as shown in Fig.~\ref{fig:beams}.

\subsection{Imaging}

We image the  atoms with a collimated, circularly polarised beam, resonant with the $|F=2, m_F=2\rangle \rightarrow |F'=3, m_{F'}=3\rangle$ transition. The imaging efficiency was calibrated to account for the deviations of the absorption cross-section from the theoretical value, primarily due to imperfections in the polarisation of the imaging light. We compared the measured atom number at the BEC critical point $\tilde N_c$ with the theoretically predicted critical number $N_c$. $\tilde N_c$ was measured by extrapolating the non-saturation slope of the thermal component to the condensation point as described in \cite{Tammuz2011, Smith2011}. We found an imaging efficiency of $(1.7 \pm 0.2)^{-1}$, which lies in the range typical of other experiments \cite{Gerbier2004,Sanner2010}.

\section{Laser Cooling}\label{sec:MOT}

For the MOT  we use an axial field gradient of $B' = 13 \unit{G/cm}$ and a cooling-light detuning of $-25\unit{MHz}$. Typically, we load the MOT to $\approx 10^9$ atoms in $15 \unit{s}$. Note, however, that $5\;$s of loading are sufficient to produce BECs with $>10^5$ atoms. In daily operation, MOT atom numbers are deduced from the intensity of fluorescence light collected on a photodiode, which was calibrated using absorption imaging. 

Following the initial loading stage, we compress the MOT by increasing $B'$ to $64\unit{G/cm}$ for
$20\unit{ms}$, and then detune the cooling light to $-68\unit{MHz}$ for $1\unit{ms}$ (CMOT, \cite{Petrich1994}). We found that these steps reduce the temperature and improve transfer into the magnetic trap.
These steps are followed by $2 \unit{ms}$ of optical molasses during which the magnetic field is switched off and the cooling-light detuning of $-68\unit{MHz}$ is maintained. This cools the cloud to $\approx15\unit{\mu K}$.

\section{Condensation in a Hybrid Trap}\label{sec:Hybrid}

Following molasses, the atoms are optically pumped into the $|F, m_F\rangle = |2,2\rangle$ hyperfine ground state, using a combination of resonant $|F=2\rangle \rightarrow |F'=2\rangle$ $\sigma^+$ light and repump light.
We pump $>80\%$ of the atoms into the $|2,2\rangle$ state in $40\unit{\mu s}$, while heating the cloud to $40\unit{\mu K}$.

The atoms are then captured in a magnetic potential by suddenly turning on the quadrupole field with $B' = 64\unit{G/cm}$ (matching the gradient of the CMOT) and then ramping $B'$ to $80\unit{G/cm}$ over $200\unit{ms}$. Finally, the trap is further compressed by raising $B'$ to $200\unit{G/cm}$ over $500\unit{ms}$.

From this point on, the evaporative cooling to condensation can be divided into three stages: (\textbf{I}) RF evaporation, (\textbf{II}) transfer into the hybrid trap, and  (\textbf{III}) ODT evaporation.

\begin{figure}[t]
\begin{center}
\includegraphics[width=\linewidth]{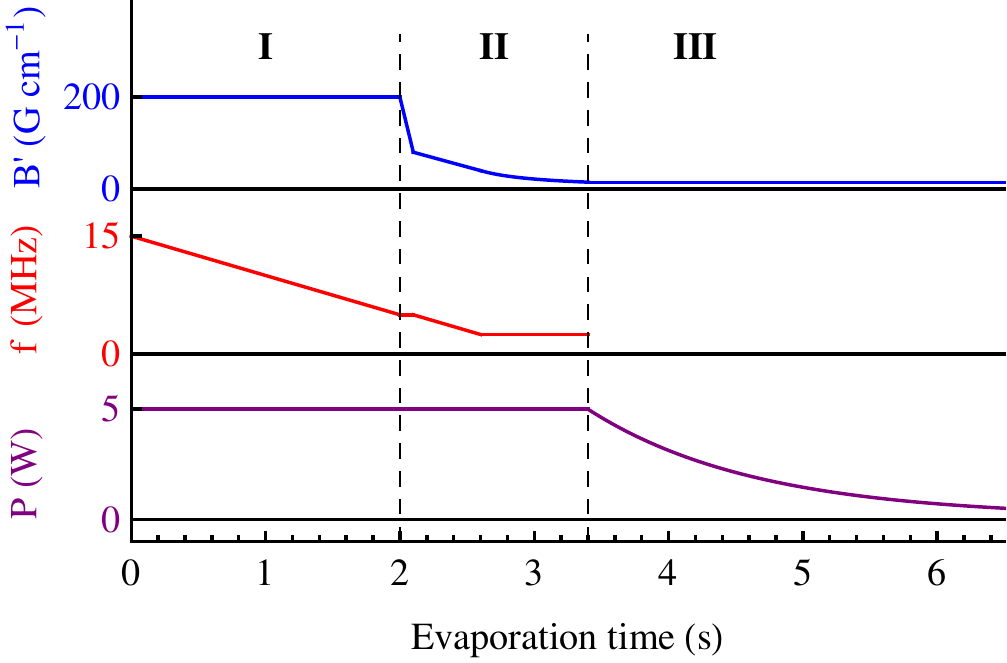}
\end{center}
\caption{Evaporation sequence in the hybrid trap. We show the evolution of (top to bottom) the magnetic field gradient $B'$, RF frequency $f$, and ODT power $P$. The sequence has three stages: (\textbf{I}) RF evaporation, (\textbf{II}) transfer into the hybrid trap, and  (\textbf{III}) ODT evaporation.}
\label{fig:sequence}
\end{figure}

\begin{figure}[t]
\begin{center}
\includegraphics[width=\linewidth]{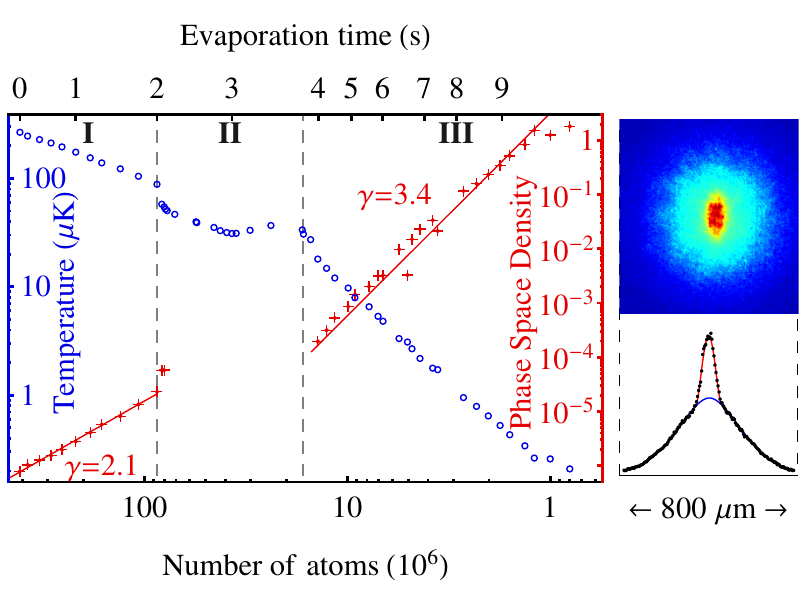}
\end{center}
\caption{BEC production. We plot the temperature $T$ (blue open circles, left axis) and phase-space density $D$ (red crosses, right axis) versus atom number $N$ (bottom) and time $t$ (top) throughout the cooling sequence. See Fig.~\ref{fig:sequence} and text for details of stages \textbf{I} - \textbf{III}. Also shown are an absorption image of a partially condensed cloud after $50\unit{ms}$ of time-of-flight expansion [optical density varies between 0 (blue) and 3 (red)] and its integrated profile. Fits to the thermal and condensed components are indicated.}
\label{fig:NT}
\end{figure}

Fig.~\ref{fig:sequence} summarises the evolution of the relevant experimental parameters during the evaporation sequence, while Fig.~\ref{fig:NT} displays the evolution of the atom number $N$, the temperature $T$, and the calculated phase-space density $D$.

$N$ and $T$ are measured using time-of-flight (TOF) absorption imaging and $D$ is calculated from a semiclassical model \cite{Lin2009}, using an analytical expression for the hybrid trapping potential. We verified the parameters of the model (such as beam waist) by measuring trapping frequencies and comparing them with theoretical predictions. 
We omit the calculated values of $D$ in stage \textbf{II}, where they are unreliable because the trapping potential varies slowly over large volumes.

\subsection{RF Evaporation}

During the initial evaporation stage in the compressed trap (\textbf{I} in Figs.~\ref{fig:sequence} and \ref{fig:NT}), we ramp the RF frequency linearly from $15\unit{MHz}$ to $5\unit{MHz}$ in $2\unit{s}$. We achieve a 30-fold increase in the peak phase space density, with an evaporation efficiency $\gamma = -{\rm d}[\ln{D}]/{\rm d}[\ln{N}]=2.1$. 

At the end of this stage the cloud typically contains $N=50\ee{6}$ atoms at $T=90\unit{\mu K}$.

\subsection{Loading of the Hybrid Trap}

In stage \textbf{II} the magnetic field gradient is decompressed to just below the gravity-compensating value, $B'_g = mg/\mu_B = 15.3\unit{G/cm}$, where $m$ is the atom mass, $g$ is the gravitational acceleration and $\mu_B$ is the Bohr magneton.
From this point on, the atoms are supported against gravity by optical forces only, while the confinement along the ODT axis is dominated by the magnetic forces~\cite{Lin2009}. Note that the ODT beam is on from the beginning of the magnetic trapping (see Fig.~\ref{fig:sequence}), but initially does not have a dominant role. 

We decompress $B'$ to $80\unit{G/cm}$ over $100\unit{ms}$ while keeping the RF frequency constant, then decompress further to $B'_1=40\unit{G/cm}$ while sweeping the RF frequency linearly to $2.5\unit{MHz}$  in $0.5\unit{s}$; this provides a good gain in phase-space density despite the weak magnetic confinement. Finally, the magnetic field is decompressed from $B'_1$ to $B'_2 = 14.8\unit{G/cm}$. In this step we sweep the field gradient according to $B'(t) = B'_1\times\left(1+t/\tau\right)^{-1}$, where $\tau = t_2 B'_2/(B'_1-B'_2)$ and the sweep duration is $t_2 = 800\unit{ms}$. This changes the RF-limited trap depth approximately linearly with time, with  the approximation being exact in the case of no ODT  and $B'_2=B'_g$.

At the end of this stage we typically have  $20\times10^6$ atoms at $30\unit{\mu K}$.

\subsection{Optical Evaporation}

For $B'<B'_g$, evaporation is no longer driven by the RF field. Instead, the trap depth is limited by the potential height at a saddle point vertically below the magnetic-field zero \cite{Hung2008a}. Evaporation can thus be forced efficiently by lowering the ODT power, $P$, albeit at the cost of a small reduction in the trapping frequencies. 

We ramp $P$ exponentially to its final value, $P_\textrm{end}$, with a time constant of $1.25\unit{s}$ and total ramp duration of $7\unit{s}$. (Both parameters have been optimised empirically.) We achieve an evaporation efficiency of $\gamma=3.4$, leading to condensation at $P_\textrm{end} = 0.16\unit{W}$.

At the critical point, $N \approx 10^6$ and $T \approx 250\unit{nK}$. At this point, the trapping frequency is about 30~Hz along the ODT axis and about 90~Hz radially.

Lowering the ODT power further results in quasi-pure condensates of $>3\ee{5}$ atoms. The lifetime of the quasi-pure BEC is $\sim3\unit{s}$, consistent with the expected 3-body losses \cite{Burt1997}.

\section{Condensation in a Plugged Quadrupole Trap}\label{sec:plug}

Our experiments with the optically plugged quadrupole trap were performed before we could introduce the isotopically pure $\Rb$ vapour sources into the setup. For this reason the atom numbers in the hybrid and plugged traps cannot be directly compared. The $28\%$ abundance of $\Rb$ in the natural isotopic mixture means that for the same background pressure, hence the same MOT loading time and cloud lifetime in the magnetic trap, the MOT loading rate is about 4 times smaller. The collisional rate, which determines the efficiency of evaporative cooling, is then also correspondingly lower.
As a result of these effects, we could produce only smaller BECs, containing up to $\sim 3 \times 10^4$ atoms. 

Our goal here, however, was not to maximise the atom number, but to investigate the minimal requirements on the plug-laser power. The improvements brought about by using isotopically pure $\Rb$ sources should equally increase the degenerate sample size in plugged-trap experiments. 

In the first plugged-trap experiment, $5\;$W of 532 nm laser power was used~\cite{Davis1995}, while later it was shown that by reducing the plug size this requirement can be reduced to $1\;$W~\cite{Naik2005}. Here we show that bringing the laser wavelength closer to the atomic resonance can reduce this value to as little as $70\;$mW.

\subsection{Cooling Sequence}

In these experiments, the laser-cooling stage was essentially the same as in Sec. \ref{sec:MOT}, except that the optimisation of the rubidium pressure for best BEC production resulted in $\sim 1.5\ee{8}$ atoms being loaded into the MOT.

The atoms were then loaded into the plugged quadrupole trap with an axial gradient of $360\unit{G/cm}$. The strong magnetic confinement compensates for the smaller initial atom number and allows the initiation of evaporative cooling.
RF evaporation started at $f = 16\unit{MHz}$. After holding $f$ constant for $1\unit{s}$, we swept it linearly to $4\unit{MHz}$ in $2\unit{s}$. At this point the cloud was sufficiently dense for the losses due to three-body recombination to become relevant~\cite{Dubessy:2012}. We thus decompressed the trap to $160\unit{G/cm}$ in $300\unit{ms}$.
The RF frequency was then swept exponentially from $3\unit{MHz}$ to $\sim180\unit{kHz}$ in $2.5\unit{s}$. 

\subsection{Effects of Plug Power and Wavelength}

\begin{figure}[t]
\begin{center}
\begin{tabular}{l}
(a) \\
\includegraphics[width=1\linewidth]{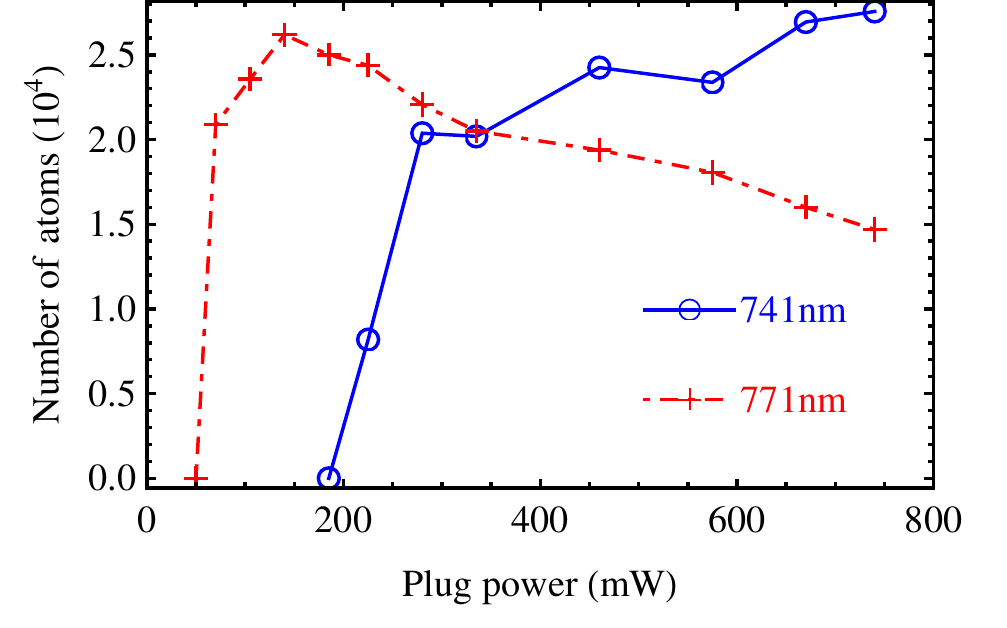}\\
(b) \\
 \includegraphics[width=1\linewidth]{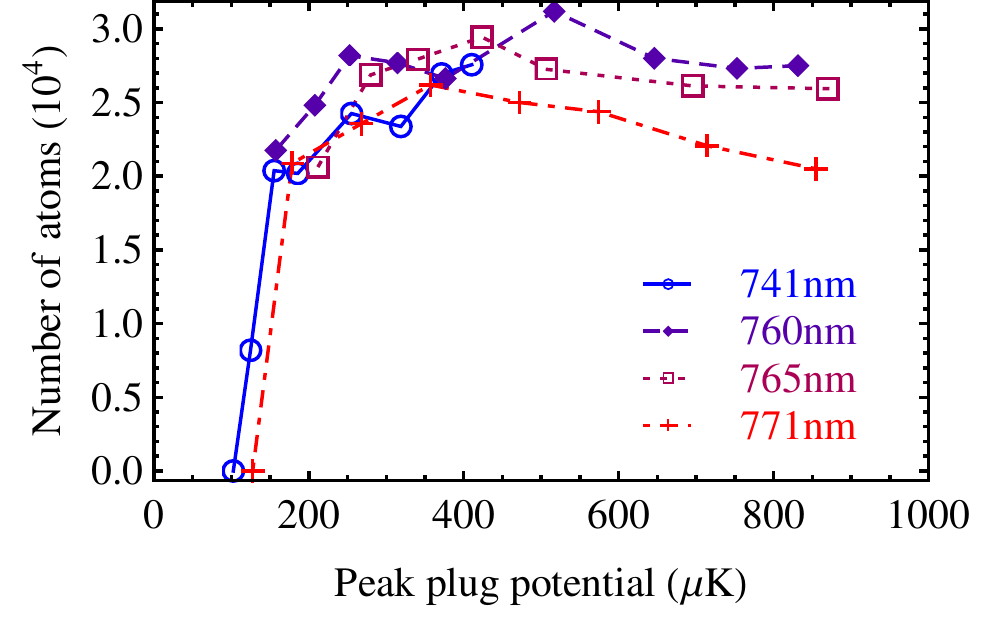}
\end{tabular}
\end{center}
\caption{Condensate atom number for different plug parameters. (a) Atom number versus plug power for two different wavelength. At 771~nm only 70~mW of plug power are necessary for condensation. (b) Atom number versus the peak potential created by the plug.}
\label{fig:plug}
\end{figure}

In the above cooling sequence, the plug laser beam was on from the moment the atoms were loaded into the quadrupole magnetic trap. However, its efficiency in preventing Majorana losses depends on its power and wavelength.
Reducing the beam detuning from the atomic resonance increases both the optical dipole forces and the spontaneous light scattering, thus allowing smaller powers to be used, but at the cost of more heating and shorter lifetimes. 

To study this effect quantitatively, we varied the plug-laser wavelength in the range $740-775\unit{nm}$ and the beam power, $\Plug$, in the range $10\;$mW - $1\unit{W}$, while keeping the beam geometry fixed. 

Fig.~\ref{fig:plug}(a) shows the number of atoms in the condensed cloud as a function of $\Plug$ for two different wavelengths.
At $741\;$nm we achieve condensation for $\Plug \gtrsim 200\;$mW; increasing $\Plug$ beyond 300~mW leads to only a slow increase in the atom number. At $771\;$nm, as little as 70~mW of plug power is sufficient for condensation. Note, however, that in this case one should use the minimum necessary laser power, since  increasing $\Plug$ further actually has adverse effects due to the increased spontaneous light scattering.
 
In Fig.~\ref{fig:plug}(b) we plot the number of atoms at the end of the evaporation versus the peak value of the repulsive potential created by the plug beam. We see that the threshold potential is in fact the same for a range of wavelengths. This also shows that (for this range of wavelengths) the spontaneous light scattering is not relevant for powers up to the threshold value. Only for higher peak potentials we see that longer wavelengths (closer to resonance) lead to lower atom numbers due to the spontaneous scattering.

\section{Conclusion}\label{sec:conclusion}
We have described a simple single-chamber apparatus for producing  Bose-Einstein condensates of $\Rb$ by evaporative cooling in either a hybrid magnetic-optical trap or an optically plugged quadrupole trap.
With the hybrid-trap approach we achieve BECs with $> 3 \times 10^5$ atoms, comparable to most multi-chamber setups.
In our experiments with the optically plugged trap the main result is that as little as 70~mW of plug power is sufficient to achieve condensation. This makes it possible to pursue this approach using inexpensive diode lasers.

We hope that our description of both experimental approaches will benefit groups wanting to design simple and cost-efficient BEC machines for various applications.

\acknowledgments
This work was supported by EPSRC (Grant No. EP/G026823/1), the Royal Society and a grant from the Army Research Office with funding from the DARPA OLE program.


\begin{thebibliography}{10}

\bibitem{Bloch:2008}
I. Bloch, J. Dalibard, and W. Zwerger, Rev. Mod. Phys. {\bf 80},  885  (2008).

\bibitem{Cronin:2009}
A.~D. Cronin, J. Schmiedmayer, and D.~E. Pritchard, Rev. Mod. Phys. {\bf 81},
  1051  (2009).
  
 \bibitem{Anderson1995}
M.~H. Anderson {\it et~al.}, Science {\bf 269},  198  (1995).

\bibitem{Bradley1995}
C.~C. Bradley, C.~A. Sackett, J.~J. Tollett, and R.~G. Hulet, Phys. Rev. Lett.
  {\bf 75},  1687  (1995).

\bibitem{Davis1995}
K.~B. Davis {\it et~al.}, Phys. Rev. Lett. {\bf 75},  3969  (1995).

\bibitem{Barrett2001}
M.~D. Barrett, J.~A. Sauer, and M.~S. Chapman, Phys. Rev. Lett. {\bf 87},
  010404  (2001).

\bibitem{Tablieber:2008}
M. Taglieber {\it et~al.}, Phys. Rev. Lett. {\bf 100},  010401  (2008).

\bibitem{Bakr:2009}
W.~S. Bakr, A. Peng, S. Folling, and M. Greiner, Nature {\bf 462},  74  (2009).

\bibitem{Zipkes:2010}
C. Zipkes, S. Palzer, C. Sias, and M. K\"ohl, Nature {\bf {464}},  388
  ({2010}).

\bibitem{Lewandowski2003}
H.~J. Lewandowski, D.~M. Harber, D.~L. Whitaker, and E.~A. Cornell, J. Low
  Temp. Phys. {\bf 132},  309  (2003).

\bibitem{Du2004}
S. Du {\it et~al.}, Phys. Rev. A {\bf 70},  053606  (2004).

\bibitem{Naik2005}
D.~S. Naik and C. Raman, Phys. Rev. A {\bf 71},  033617  (2005).

\bibitem{Heo2011}
M.-S. Heo, J.-Y. Choi, and Y.-I. Shin, Phys. Rev. A {\bf 83},  013622  (2011).

\bibitem{Weber2003}
T. Weber {\it et~al.}, Science {\bf 299},  232  (2003).

\bibitem{Kinoshita2005}
T. Kinoshita, T. Wenger, and D.~S. Weiss, Phys. Rev. A {\bf 71},  011602
  (2005).

\bibitem{Jacob2011}
D. Jacob {\it et~al.}, New J. Phys. {\bf 13},  065022  (2011).

\bibitem{Arnold2011a}
K. Arnold and M. Barrett, Opt. Commun. {\bf 284},  3288  (2011).

\bibitem{Lin2009}
Y.-J. Lin {\it et~al.}, Phys. Rev. A {\bf 79},  063631  (2009).

\bibitem{Gaunt:2013}
A. Gaunt {\it et al.}, arXiv:1212.4453 (2012).

\bibitem{Lauber2011b}
T. Lauber, J. K\"uber, O. Wille, and G. Birkl, Phys. Rev. A {\bf 84},  043641
  (2011).

\bibitem{Tammuz2011}
N. Tammuz {\it et~al.}, Phys. Rev. Lett. {\bf 106},  230401  (2011).

\bibitem{Smith2011}
R. P. Smith, R. L. D. Campbell, N. Tammuz, and Z. Hadzibabic,
Phys. Rev. Lett. {\bf 106}, 250403 (2011).

\bibitem{Gerbier2004}
F. Gerbier {\it et~al.}, Phys. Rev. Lett. {\bf 92},  030405  (2004).

\bibitem{Sanner2010}
C. Sanner {\it et~al.}, Phys. Rev. Lett. {\bf 105},  040402  (2010).

\bibitem{Petrich1994}
W. Petrich, M.~H. Anderson, J.~R. Ensher, and E.~A. Cornell, J. Opt. Soc. Am. B
  {\bf 11},  1332  (1994).

\bibitem{Hung2008a}
C.-L. Hung, X. Zhang, N. Gemelke, and C. Chin, Phys. Rev. A {\bf 78},  011604
  (2008).

\bibitem{Burt1997}
E.~A. Burt {\it et~al.}, Phys. Rev. Lett. {\bf 79},  337  (1997).

\bibitem{Dubessy:2012}
R. Dubessy {\it et al.}, Phys. Rev. A {\bf 85}, 013643 (2012). 

\end{thebibliography}

\end{document}